\documentclass[12pt,aps,amsmath,latexsym,amsfonts,letterpaper]{JHEP3}

\usepackage{graphicx}
\usepackage{epsfig}
\usepackage{amssymb}
\usepackage[all]{xy}
\usepackage{subfig}
\usepackage{amsfonts}

\def\half{\textstyle{1\over2}}

\def\nn{\nonumber}

\newcommand{\be}{\begin{equation}}
\newcommand{\ee}{\end{equation}}
\newcommand{\bea}{\begin{eqnarray}}
\newcommand{\eea}{\end{eqnarray}}

\newcommand{\lab}{\label}

\title{The gravitational dynamics of warped throats}
\author{Neil A. Butcher\footnote{email: ppxnb@nottingham.ac.uk}  and 
        Paul M. Saffin\footnote{email: paul.saffin@nottingham.ac.uk}\\
School of Physics and Astronomy, University Park, University of
Nottingham, Nottingham NG7 2RD, UK}

\date{\today}

\maketitle

\abstract{
We investigate the time evolution due to gravitational dynamics of a particular spacetime commonly used in brane-cosmology and string compactifications, namely the Klebanov-Strassler geometry, which is achieved by adding a perturbation to the momentum of the static solution. We observe the effects this has on the spacetime and look for evidence of black hole formation or collapsing cycles which could lead to singular geometry. The cycles are seen to commonly re-expand after reaching a minimum value, showing the stability of the solution against perturbations which would change its size. However black holes are observed to form for certain perturbations, which could impede common uses of the throat's stable tip.

 }

\keywords{\it compactification, topology change, conical singularity, black hole.}
\preprint{arXiv:yymm.nnnn [hep-th]}

\begin{document}

\section{Introduction} 
\lab{intro}

Producing realistic models of our own universe using the knowledge we have of string theory is a task plagued with difficulties, such as the selection of a vacuum out of many choices, and the existence of moduli. String theory requires that extra dimensions are present, but observations restrict them to be unobserved. Such extra dimensions may be in the form of a compact manifold, the choice of which is a choice of vacuum with the moduli being continuous parameters characterizing the manifold. These moduli take the form of massless scalar fields within the low energy effective theory,  which are strongly constrained\cite{deCarlos:1993jw,Dine:1985he}. With appropriate addition of fluxes and the inclusion of quantum mechanical effects these moduli fields can take on an expectation value and at the same time can acquire a mass, this is desirable as a large mass would explain why these fields have not been detected. In addition to the values taken by the moduli is the more fundamental question of the topology of the compact manifold, however the seemingly insurmountable challenge of selecting one topology was improved by the possibility that vacua with different topology are connected by paths in the moduli space, ones going through singular geometry manifolds\cite{Greene:1996cy,Candelas:1988di,Candelas:1989ug,Green:1988bp,Green:1988wa}. Such a change to the topology could be a very drastic process, altering intersection numbers or even the Hodge numbers of the compact topology. These transitions can be made regular, and their low energy dynamics studied within the realm of string theory\cite{Strominger:1995cz,Greene:1996dh,Lukas:2004du,Palti:2005kv}.
The range of possible vacuum topology and moduli is collectively called the string landscape\cite{Susskind:2003kw} and our own position within this huge range of vacua will determine many of the phenomenological predictions of the construction. With the introduction of an expectation value to the moduli fields some points within the string landscape become preferable and the universe will flow to these points, maybe even changing topology to get there if need be. With this additional knowledge about the Calabi Yau manifold being compactified on, string theory can make phemomonological predictions which may test it. 

Warped throats are a possible feature within flux compactifications, they look like an extended protrusion of the manifold, whose base does not grow in size very quickly as we move away from the tip. Previously extra dimensions have been compactified upon manifolds with such a region in efforts to produce inflationary models (warm inflation, brane inflation, DBI inflation, spinflation)
;  recreating the standard model using anti-D3 branes\cite{Cascales:2003wn}
; creating a hierarchy between the UV compactification scale and the IR at the tip of the throat\cite{Giddings:2001yu}
or other interesting phenomenological effects. These generally involve using this manifold as the background upon which some probe brane is moved, where it is assumed that the probe does not influence the background manifold. Warped throats are often chosen as a background manifold also because they possess only a few moduli with flat directions. Most parameters which define the size, shape and other properties of the throat, such as the dilaton and the complex structure moduli, may be stabilized by the choice of flux, making the moduli at the supersymmetric vacua (potential minima) precisely determined. The stabilized moduli are less susceptible to backreaction from the branes and other probes added to create the wanted interesting effects like inflation.

A good example of such a throat is the deformed conifold which can be combined with fluxes to give a Klebanov-Strassler throat\cite{Klebanov:2000hb}, this solution is the one commonly used in investigations involving warped throats; it is this solution which we too will use as a starting condition as we proceed to model the evolution of warped throats in time. We intend to use type IIB supergravity to perform a numerical simulation of the gravitational effects of perturbing the throat in the region around its tip. Unlike some previous studies, generally performed within the 4D-effective low energy theory of the moduli space, we do not trust the moduli approximations alone due to the immense freedom of the flux-containing, gravitationally-backreacting spacetime of IIB supergravity. We hope that our more elaborate numerical simulations (ones able to detect black hole formation and effects undetectable in the moduli approximation) will shed more insight on the possible limitations of these previous attempts.

\section{Lagrangian and equations of motion}
Our study takes place within the framework of type IIB supergravity, whose action is given in the Einstein frame as follows \cite{Polchinski:1998rr,Cvetic:2000gj},
\bea
\mathcal{L}_{10}^{IIB}=&&R*\mathbb{I}+\half d\phi \wedge *d\phi-\half e^{2\phi} F_1\wedge *F_1 -\frac{1}{4} F_5\wedge *F_5\nn\\&& -\half e^{-\phi} H\wedge *H -\half e^{\phi} F_3\wedge *F_3-\half C_4 \wedge H \wedge F_3
\eea
The fields comprise a scalar dilaton $\phi$, an R-R scalar with field strength $F_1=dC_0$, an R-R two form with field strength $F_3$, an NS-NS three form field strength $H=dB$ and a self dual five form field strength $F_5=dC_4+B \wedge (F_3+C_0\:H)$. We have made gauge choices as we write these potentials, even though the four potential, $C_4$, appears in the Chern-Simmons term of the Lagrangian it still has a great deal of gauge freedom.  We need to separately impose the self duality condition upon $F_5$ by hand, as it is unconstrained by the Lagrangian
\be
F_5=*F_5.
\ee
The Lagrangian gives the following equations of motion for these fields:
\bea
d(*F_5)&=&-F_3\wedge H\\
d(e^{\phi}\,*F_3)&=&F_5 \wedge H\\
d(e^{2\phi}\,*F_1)&=&e^{\phi}\:*F_3 \wedge H\\
d(e^{-\phi}\,*H)&=&e^{\phi}\:F_1 \wedge *F_3-F_5 \wedge F_3\\
d*d \phi&=&-e^{2\phi}\,*F_1\wedge F_1-\half e^{\phi}\,*F_3\wedge F_3+\half e^{-\phi}\,*H\wedge H
\eea
The fluxes also contribute to the energy momentum tensor, which means that our spacetime will not be Ricci flat but will obey the Einstein equation,

%

 \bea
 R_{MN}&=&\half \partial_M\,\phi\,\partial_N\,\phi+\half\,e^{2\phi}\,F_{(1)}\,_{M}F_{(1)}\,_{N}+\frac{1}{96}\,F_{(5)}\,_M\,^{abcd}F_{(5)}\,_{Nabcd}\nn\\
 &&+\frac{1}{4}e^{+\phi}\,\left(\,F_{(3)}\,_{M}\,^{ab}F_{(3)}\,_{Nab}-\frac{1}{12}F_{(3)}\,^2\,g_{MN}\right)\\
 &&+\frac{1}{4}e^{-\phi}\,\left(\,H\,_{M}\,^{ab}H\,_{Nab}-\frac{1}{12}H\,^2\,g_{MN}\right)\nn.
 \eea

In the above we have been using a mostly plus metric signature and an antisymmetric tensor obeying
\be
\epsilon_{0\,1\,2\,3...}=+1,
\ee
with the ten dimensional Hodge dual being given by
\be
*e^{abc..}=\frac{1}{n!}\,\epsilon^{abc..}\,_{i_1 i_2..i_n}e^{i_1 i_2..i_n}
\ee
\section{Klebanov-Strassler static solution}
\lab{Kleb}
A static warped throat solution found by Klebanov and Strassler \cite{Klebanov:2000hb} will be used as our starting condition to which we will add a small perturbation in the form of a momentum. This in turn was based on the deformed conifold \cite{Candelas:1989js}, which is a Calabi-Yau space with one extended radial dimension, labelled by $r$, and a compact five dimensional base. At high $r$, far from the origin, the deformed conifold tends to look like the standard conifold (where the five dimensional base is a Sasaki Einstein metric with the topology $S^2$x$S^3$). However at lower radius the deformed conifold has a base dependent on r, and so loses scale invariance in $r$, the new scale being introduced by the new parameter $\epsilon$.
We can write this line element in terms of a basis of one forms, $g^i$ defined and described in detail in\cite{Klebanov:2000hb}. Using this basis of one forms to describe the five dimensional base of $ds_6$, we can write the deformed conifold as
\bea
ds_6=&\half&\epsilon^{\frac{4}{3}}\,K(r)\:\left(\frac{1}{3K^3(r)}\right)dr^2\nn\\
+&\half&\epsilon^{\frac{4}{3}}\,K(r)\:\left(sinh^2\left(\frac{r}{2}\right)\right) ((g^1)\,^2+(g^2)\,^2)\nn\\
+&\half&\epsilon^{\frac{4}{3}}\,K(r)\:\left(cosh^2\left(\frac{r}{2}\right)\right) ((g^3)\,^2+(g^4)\,^2)\nn\\
+&\half&\epsilon^{\frac{4}{3}}\,K(r)\:\left(\frac{1}{3K^3(r)}\right)((g^5)\,^2).
\eea
Where we have used
\be
\label{eq:K(r)}
K(r)=\frac{\left(sinh(2r))-2r\right)^{\frac{1}{3}}}{2^{\frac{1}{3}} sinh(r)}.
\ee
Note that $K(r)$ does not vanish at the origin but tends to the constant value $(2/3)^{1/3}$, implying that the deformed conifold does not tend to a singular point at $r=0$, but at the tip there is a three sphere in the direction $g^3 \wedge g^4 \wedge g^5$.
This six dimensional manifold will make up six of the ten dimensions in the solution to the Einstein equations. In the case with no flux contributions we can product the deformed conifold with a 3+1 Minkowski metric $ds_{1,3}$.
\be
\lab{eq:minkmet}
ds_{10}^2=ds_{1,3}+ds_6^2.
\ee

\subsection{D-brane sourced fluxes}
If we introduce $N$ D3branes and further supplement this with $M$ D5branes wraped arround the two sphere of a deformed conifold (these have previously been coined "fractional D3branes") then it corresponds to taking a Calabi Yau manifold with deformed conifold singularity and introducing a large flux through the three sphere\cite{Klebanov:1999tb,Klebanov:2000nc}. The effect of combining $ds_6$ with (fractional) Dbranes is to source fluxes and so contribute to the Energy-Momentum tensor of the Einstein equation.
This means that the spacetime metric can no longer be Ricci flat and it changes the static solution to the solution of \cite{Klebanov:2000hb}.
The new fluxes require a change to the metric (\ref{eq:minkmet}), making the minkowski slices depend upon the radius of the deformed conifold component and so we get conformal symmetry breaking (for non-zero $M$). This $r$ dependence is introduced by means of a function, $h(r)$. 
\be
\label{eq:KlebMetric}
ds_{10}^2=h^{-\half}(r)\left(ds_{1,3}\right)+h^{\half}(r)\left(ds_6^2\right)
\ee
The static solution is described in detail in \cite{Klebanov:2000hb}, we intend to use this metric and these fluxes as initial conditions which we will then go on to perturb. The fluxes, three-forms and the five-forms of the static case are described in terms of the metric coordinates. The scalar dilaton and R-R scalar go as,

\bea
F_1&=&0,\\
e^{\phi}&=&g_{string}.\label{eq:initDil}
\eea
In addition to the usual potential, the M fractional branes we have placed at the tip of the deformed conifold, consistently give flux through the $g^5 \wedge g^3 \wedge g^4$ direction at the origin, this is the 3 sphere,
\bea
C_2&=&M \frac{\sinh(r)-r}{2 \sinh(r)}\,\left(g^1 \wedge g^3+g^2 \wedge g^4\right),\nn\\
F_3&=&M\, g^5 \wedge g^3 \wedge g^4+d C_2.\label{eq:initF3}
\eea
H can be described by the potential $B$,
\bea
B_{\alpha}&=&g_{string} M \frac{r\,\coth(r)-1}{2 \sinh(r)}\,(\cosh(r)-1),\nn\\
B_{\beta}&=&g_{string} M \frac{r\,\coth(r)-1}{2 \sinh(r)}\,(\cosh(r)+1),\nn\\
B&=&B_{\alpha} g^1 \wedge g^2 + B_{\beta} g^3 \wedge g^4,\nn\\\label{eq:initH}
H&=&dB.
\eea
The final flux is the five form, self dual $F_5$. Its self duality condition along with the equation of motion and our choices of $F_3$ and $H$ lead to the solution that,
\be
\label{eq:5flux}
F_5=B \wedge F_3 + *(B \wedge F_3).
\ee
The nature of $h(r)$ in the static case (this is the value for $h(r)$ which we impose as an initial condition) is given by a differential equation.
\be
\label{eq:h(r)}\frac{d\,h(r)}{dr}=-\alpha\, \frac{2^{\frac{2}{3}}}{4}\,\frac{r\,\coth(r)-1}{\sinh^2(r)}\,\left(\sinh(2r)-2r\right)^{\frac{1}{3}}.
\ee
The introduction of this $h(r)$ changes the scales and removes the conformal invariance. 
To totally define $h(r)$ we also need a boundary condition (to complement the 1st order differential equation), Klebanov and Strassler impose the restriction that h must vanish at high r.
\be
 \lim_{r\to\infty} h(r)=0.
\ee
Meaning that the radius of the throat tends to grow only very slowly for large r.
\subsection{Superpotential and stabilized moduli}

\FIGURE{
  \includegraphics[width=12cm]{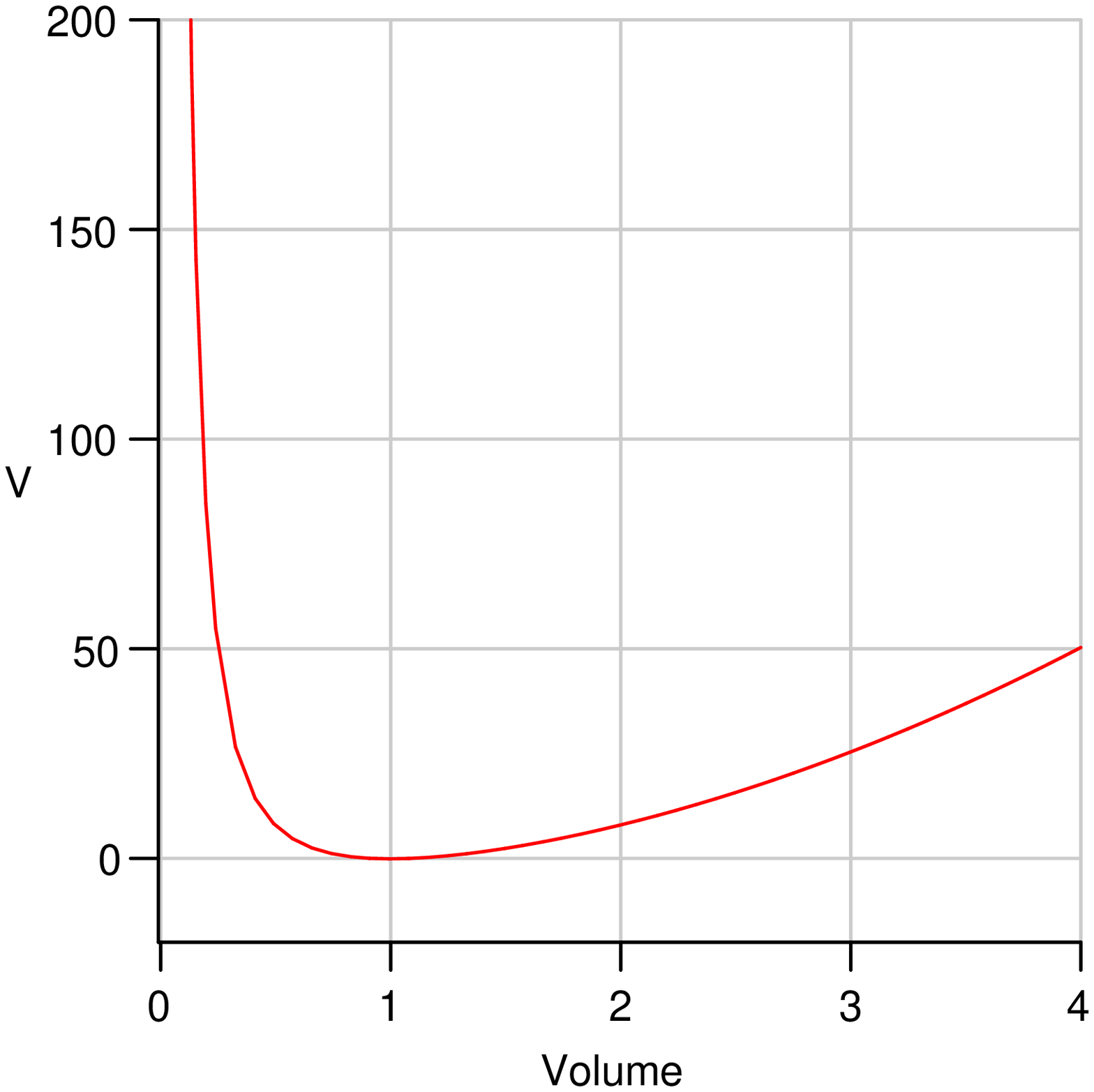}
 \caption{The potential as a function of the volume of the three sphere in planck units} 
 \label{fig:potential}}

The potential created by the introduction of the flux can be found from the Hamiltonian constraint, as described in \cite{Misner:1974qy} and used in\cite{Graham:1991av}. This potential is then used to find a prediction to the evolution of the moduli field. Taking a slice through this moduli space along which only the volume of the three-sphere is permitted to change, we find the potential as a function of this volume.
This method results in a potential of the form plotted in figure.\ref{fig:potential}.
This diverges very quickly as the volume gets small, it permits a minima at a position determined by the other moduli and parameters (such as M), and tends to grow (albeit quite slowly) as the volume gets very large. The minima represents the static warped deformed throat solution.
The scale of the compactified dimensions is now set by the fluxes, the size of the three sphere at the origin is stabilized according to the minima of the potential and is no longer a free modulus, in the static case it must conform to 
\be
\label{eq:modStab}
\epsilon=\left(\frac{16 M^2 g_{string}}{\alpha}\right)^{3/8}.
\ee

\section{Time dependent spacetime anzatz}
If we start with the static solution then no change will happen as we move forward time, however with only a small initial perturbation the metric and the fluxes are observed to evolve. It is our intention to introduce an initial perturbation that changes the size of the three sphere at the origin of the deformed conifold, possible outcomes include the formation of a black hole solution; the collapse of the three sphere to a naked singularity; or the sphere may change size without collapsing all the way to zero. To observe the effects of an initial deformation we use a more general metric and flux ansatz, a system with the capacity to be time dependent, and then observe the effects we can incite with a small initial perturbation. 

We choose a metric that is capable of changing in time and is also stable to evolve numerically at the origin\cite{Butcher:2008vc,Butcher:2007zk}.
\bea
\label{eq:metric}
ds_{10}^2=&&T(t,r)^2\: h^{-\half}(r) \left(ds_{1,3}\right)\nn\\
&&+a^2(t,r)\: h^{\half}(r)\,\half\epsilon^{\frac{4}{3}}\,K(r)\:\left(\frac{1}{3K^3(r)}\right) dr^2\nn\\
&&+b^2(t,r)\: h^{\half}(r)\,\half\epsilon^{\frac{4}{3}}\,K(r)\:\left(sinh^2(r/2)\right) ((g^1)\,^2+(g^2)\,^2)\nn\\
&&+c^2(t,r)\: h^{\half}(r)\,\half\epsilon^{\frac{4}{3}}\,K(r)\:\left(cosh^2(r/2)\right) ((g^3)\,^2+(g^4)\,^2)\nn\\
&&+d^2(t,r)\: h^{\half}(r)\,\half\epsilon^{\frac{4}{3}}\,K(r)\:\left(\frac{1}{3K^3(r)}\right) ((g^5)\,^2).
\eea
The profiles $T(t,r)$,$a(t,r)$,$b(t,r)$,$c(t,r)$ and $d(t,r)$ define the metric at all times. We also had to impose boundary conditions at the origin of the simulation. These conditions were to ensure that local flatness remained at later times\cite{Alcubierre:2004gn}.
\bea
\lab{eq:locFlatcd} c^2(t,r)|_{r=0}=d^2(t,r)|_{r=0},\\b^2(t,r)|_{r=0}=a^2(t,r)|_{r=0} \lab{eq:locFlatab}.
\eea
We also required that all these profile functions were always even at the origin.
At later times the size of the three sphere can be found from $d^2(t,r)|_{r=0}$ and $c^2(t,r)|_{r=0}$.

Of course we must also allow the fluxes to change with time (as they almost certainly will when the metric is perturbed). 
Initially the axion is constant and the equations of motion show this can continue to be the case at later times, also the dilaton is initially constant at all points however this is permitted to change at later times according to the equations of motion,
\bea
F_1&=&0,\\
\phi&=&\phi(t,r).
\eea
The M fractional branes that we have placed at the origin will not change but will always give flux through the three-sphere, however the potential $C_2$ can change,
\bea
C_2&=&C_{\alpha}(t,r)\,\left(g^1 \wedge g^3+g^2 \wedge g^4\right),\nn\\
F_3&=&M\, g^5 \wedge g^3 \wedge g^4+d C_2.
\eea
The other three-form flux, $H$, is described by two separate functions $B_{\alpha}(t,r)$ and $B_{\beta}(t,r)$, these are used as a description of $B$, and $B$ defines $H$.
\bea
B&=&B_{\alpha}(t,r) g^1 \wedge g^2 + B_{\beta}(t,r) g^3 \wedge g^4,\nn\\
H&=&dB.
\eea
Even at later times the five-form flux is still determined by the other fluxes according to (\ref{eq:5flux}).

All the fluxes are defined by the metric and four profile functions, $\phi(t,r)$, $C_{\alpha}(t,r)$, $B_{\alpha}(t,r)$ and $B_{\beta}(t,r)$. It is these functions that we will evolve using the equations of motion. In addition to the equations of motion, we also imposed boundary conditions on these functions, we required that $B_{\alpha}(t,r)$ and $B_{\beta}(t,r)$ be odd, $\phi(t,r)$ be even and $C_{\alpha}(t,r)$ be even and must vanish at the origin. This is the choice of fluxes that are capable of acting as the initial conditions and evolving consistently to later times.

\subsection{Initial conditions}
\label{sec:init}
By comparing the static Warped deformed conifold metric (\ref{eq:KlebMetric}), and our time dependent ansatz (\ref{eq:metric}), we can read off the initial metric conditions a term at a time.
\be
T^2(0,r)=
a^2(0,r)=
b^2(0,r)=
c^2(0,r)=
d^2(0,r)=1
\ee
Flux is added when we give $M$ a non-zero value (we introduce fractional branes), its strength depends upon our string coupling $g_{string}$ and the number of branes $M$. 
The initial values of the fluxes can be found from (\ref{eq:initDil}), (\ref{eq:initF3}) and (\ref{eq:initH}). They are defined by the functions
\bea
e^{\phi(0,r)}&=&g_{string},\\
B_{\alpha}(0,r)&=&g_{string} M \frac{r\,\coth(r)-1}{2 \sinh(r)}\,(\cosh(r)-1),\\
B_{\beta}(0,r)&=&g_{string} M \frac{r\,\coth(r)-1}{2 \sinh(r)}\,(\cosh(r)+1),\\
C_{\alpha}(0,r)&=&M \frac{\sinh(r)-r}{2 \sinh(r)}.
\eea

This also requires that $h(r)$ to take on the value which obeys the differential equation (\ref{eq:h(r)}) and also tends to zero asymptotically. We found $h(r)$ numerically as we input the initial conditions. 

These initial conditions give the static solution, so if all the momenta (e.g. $\dot{T}$) are zero to begin with then no evolution should occur. If instead we start with non zero momenta we perturb the metric away from the static case and can go on to see the future evolution. In order to best represent a physical system make our perturbation vanish as we go away to large distances, representing a localized perturbation. Our initial momentum must also conform to the constraints upon the Hamiltonian and the momentum imposed by the Einstein equations.
\section{Results}
\lab{results}
We kept the values of the string coupling and the number of fractional branes consistent throughout all plotted simulations, $M g_{string}=120$ and $M=30$, we also specified the warping $\alpha$ so that the static solution was at $\epsilon=1$.
If we add momentum going like (\ref{eq:mom1})-(\ref{eq:mom3}) 
\bea
\label{eq:mom1}
\frac{\dot{c}}{c}=\frac{\dot{d}}{d}&=&\: -Pe^{-r^2}\\
\label{eq:mom2}
\frac{\dot{T}}{T}\: &=&\: Pe^{-r^2}\\
\dot{a}|_{r=0}&>&0\: \: \:\: \: \:\: \: \:\: \: \:\dot{b}|_{r=0}>0
\label{eq:mom3}
\eea
Then a positive value of P will cause the size of the three sphere at the origin to initially fall but this may only be temporary, whereas a negative P will cause the three sphere to grow (the symmetry between c and d maintained the local flatness (\ref{eq:locFlatcd})). As we impose the initial conditions we must obey the Hamiltonian and momentum constraints, this requirement was used to numerically find the initial values of $\dot{a}$ and $\dot{b}$. The choice (\ref{eq:mom2}) was made to aid this numerical integration.

In order to best summarise the results of our perturbed evolution we shall constantly be watching the size of the three sphere at the origin. If this shrinks it shows that the origin is becoming closer to that of a conifold, approaching the formation of a conical singularity, with the three-sphere vanishing being the most extreme case. Alternatively we may find other outcomes, such as the formation of black holes.
\subsection{Formation of black holes and apparent horizons}
We will attempt to discover if and when black holes have formed by constantly looking for apparent horizons on the timeslice. The existence of an apparent horizon will show that there exists an event horizon outside it or coinciding with it\cite{Hawking:1973uf}. The event horizon is a sure sign of a black hole spacetime. The apparent horizon can be detected upon any single timeslice\cite{Thornburg:1995cp}, upon a timeslice with unit normal $n^i$ and where $K_{ij}$ is the extrinsic curvature of the slice, the apparent horizon is at the outermost shell of points satisfying
\be
\label{eq:BHhor}
0=\nabla _i n^i + K_{ij}n^in^j-K.
\ee
Where $K$ is the trace of $K_{ij}$. Such a shell will show that the origin is now encased within a black hole event horizon. If the area of the apparent horizon converges upon a constant value then we can take this value to be a good estimate to the area of the resultant event horizon\cite{Hawking:1973uf}.
\begin{figure}
 \subfloat[$P=50$]{\includegraphics[width=7cm]{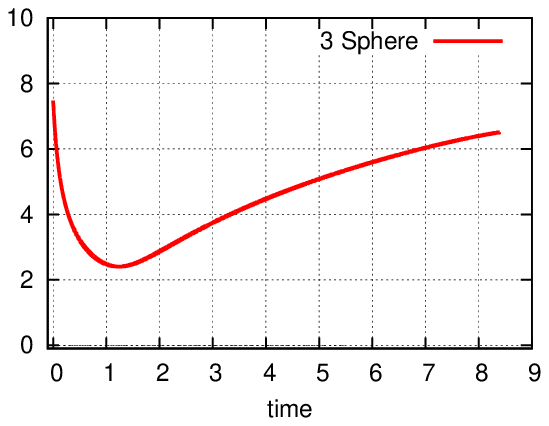}}
 \subfloat[$P=200$]{\includegraphics[width=7cm]{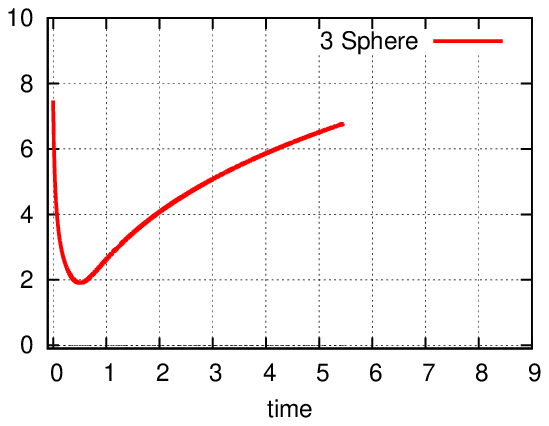}}\\
 \subfloat[$P=900$]{\includegraphics[width=7cm]{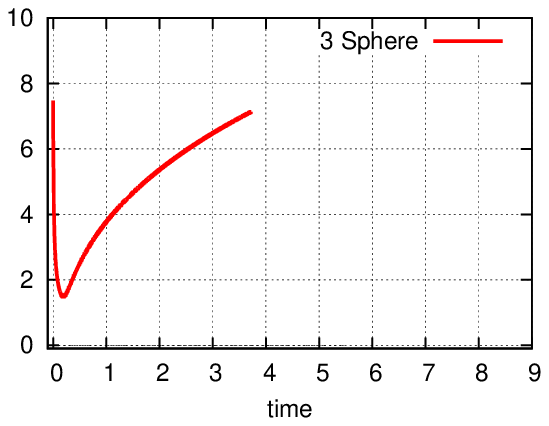}}
 \subfloat[$P=2000$]{\includegraphics[width=7cm]{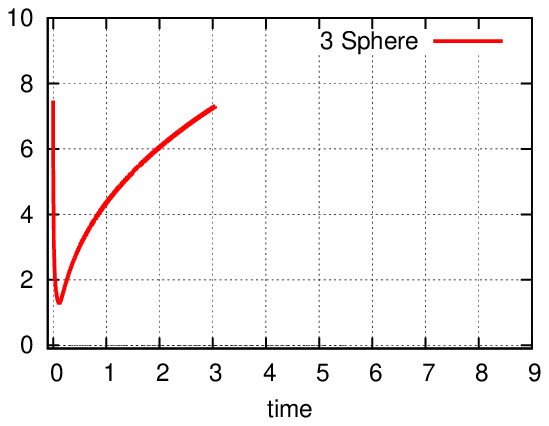}}
\caption{The size of the three sphere begins to fall but reaches a minima then returns to its starting value} 
\label{fig:bounce}
\end{figure}

\subsection{Bounce}

In order to prompt the size of the three sphere to drop we introduce a momentum of the form (\ref{eq:mom1})-(\ref{eq:mom3}) where $P>0$, this drop in the size of the sphere is, however, only temporary.
As is seen in Fig. \ref{fig:bounce}, after quickly reaching some minimum value (which depends on the strength of the momentum) the size of the three sphere then proceeds to grow, tending back to a value close to its starting value. This is an expected behaviour since the size of the three sphere is no longer a free modulus, it is determined, in the static case, by the fluxes. Since the string coupling and the number of branes is unchanged by the momentum, the ground state of the three sphere is the static value, the three sphere will tend to return to this value. In these cases no horizon is formed and the size of the three sphere tends to flow to the flux-preferred value. This can be seen to be true and quickly realized even for initial momenta hundreds of times the warped deformed scale, showing the restoring force to be very strong indeed. This is expected behaviour due to the swift divergence of the potential at low radius, as shown Fig. \ref{fig:potential}.


\subsubsection{How low can it go?}

\FIGURE{
 \includegraphics[width=12cm]{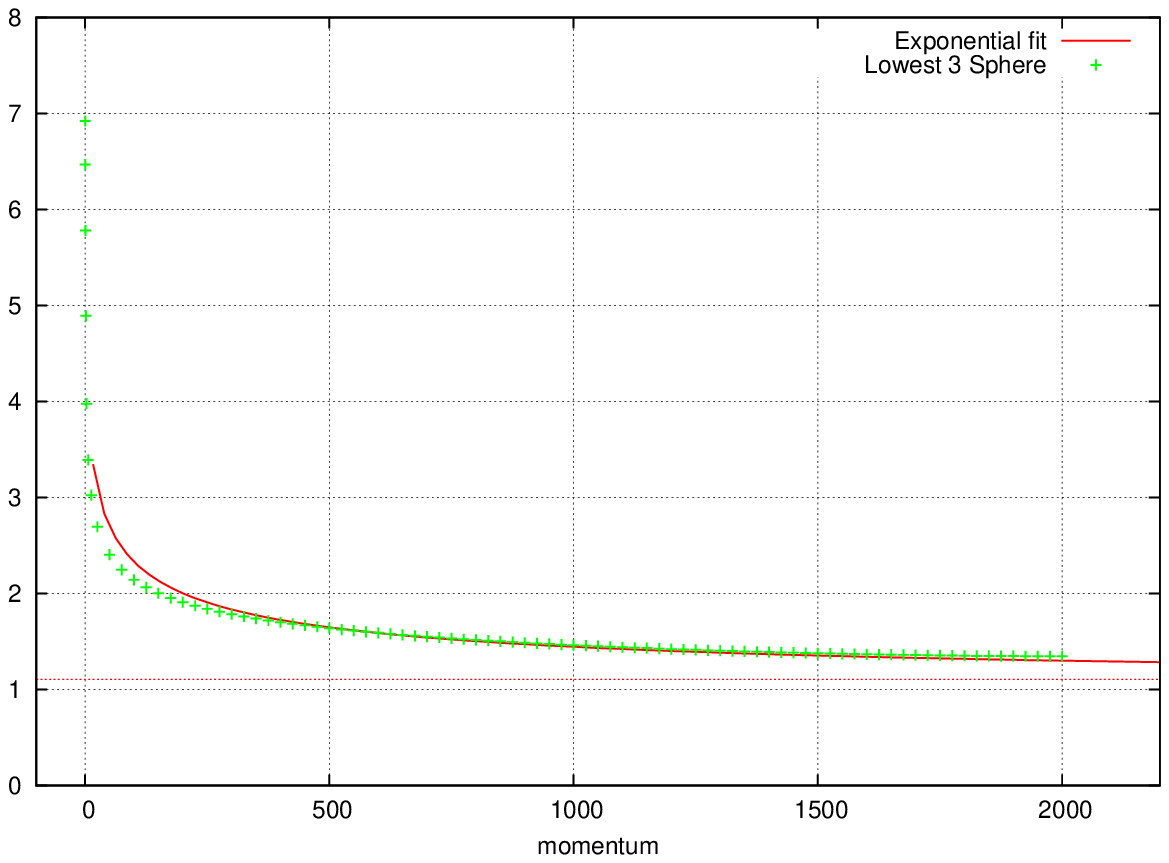}
\caption{The smallest size the three sphere reaches for a range of momenta} 
\label{fig:Lowest}}

Though the size of the three sphere can be seen to return to its initial value, it first drops to a minimum value dependent upon the initial momentum. If we continue to increase the scale of initial momentum we can ask how low we can force the three sphere to drop, could it be that there is some (very high) momentum which causes the sphere to drop to zero before it stops falling?
We can find the lowest value which the three sphere falls to for a range of initial momentum. As shown in Fig. \ref{fig:Lowest} the size does drop with the initial momentum but it drops at a decreasing rate and it would take a huge momenta to even approach zero (it actually looks as though the asymptotic behaviour may not be to zero but to a constant, lowest possible, sphere size). If we fit this to an exponential function of the form,
\be
\label{eq:fitCurve}
\alpha_0+\beta_0 e^{-\gamma_0 P^{1/4}}.
\ee
(also plotted in Fig. \ref{fig:Lowest}) then we can see that causing the sphere to vanish (if it is possible) would require incredible initial momentum way beyond the capabilities of our simulation.


\subsection{Growth}

\FIGURE{
 \includegraphics[width=12cm]{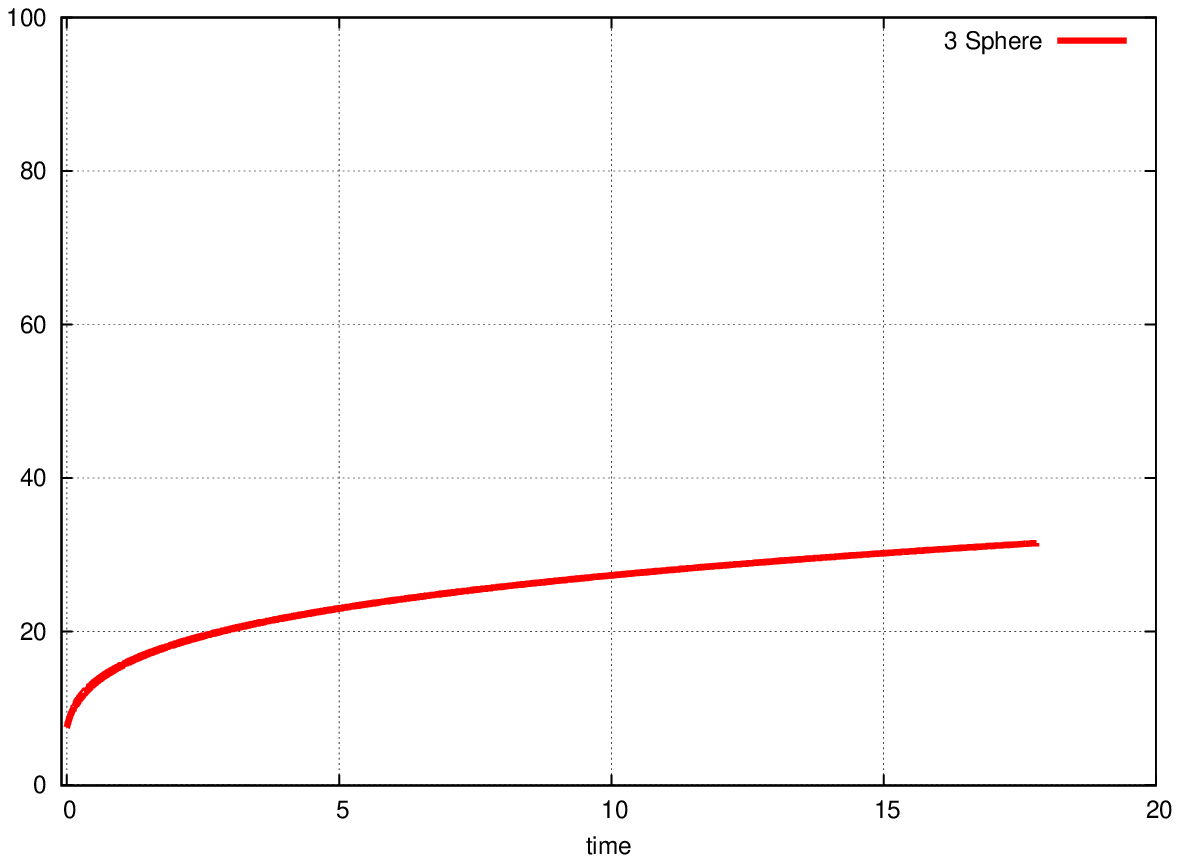}
\caption{$P=-50$: The size of the three sphere begins to grow but at a decelerating rate}
\label{fig:growth}}

We also consider cases of (\ref{eq:mom1})-(\ref{eq:mom3}) where $P<0$, these will tend to cause the size of the sphere at the origin to grow. Again we would expect (from our potential) this growth to be only temporary and that the size would fall back towards the starting value, as the static case is still determined by the fluxes and it is this value we would expect the size of the tip to flow to. We do see this slowing of the growth in Fig. \ref{fig:growth}, but slowing down takes so long that the restoration of the size is not seen within the timescaleof the simulation, this can be attributed to a shallow restoring potential. We believe that the three sphere would eventually return to the starting value (the static case) but this takes a very long time.

\subsection{Black hole formation}

\FIGURE{
 \includegraphics[width=12cm]{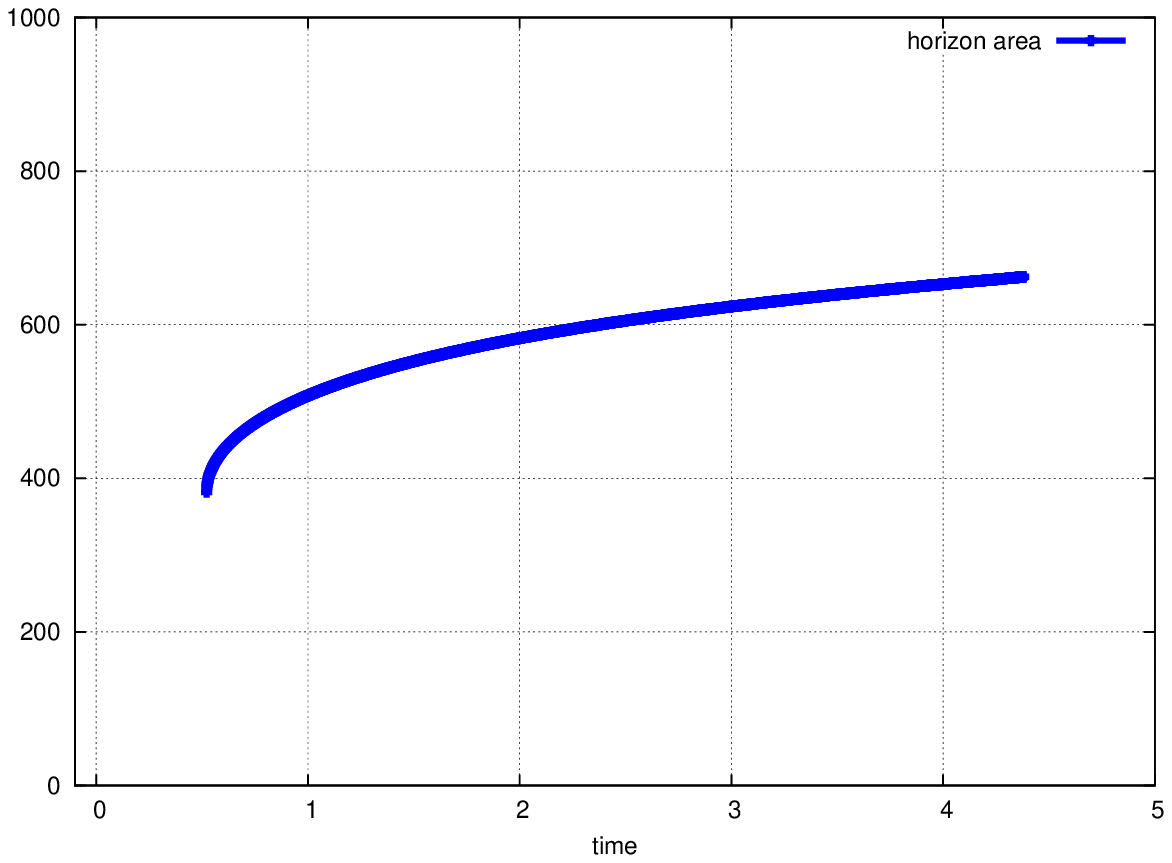}
\caption{$P=0.7$: The size of the horizon area begins to grow but at a decelerating rate}
\label{fig:horizon}}

Relaxing the condition (\ref{eq:mom3}) was also attempted. This made the situation far more susceptible to the formation of black hole horizons, detectable by shells obeying (\ref{eq:BHhor}). The presence of an apparent horizon often occurs already in the initial conditions, but the late time creation and growth of an apparent horizon is also possible, as shown in the example of Fig. \ref{fig:horizon}. These horizons would be intolerable if we wished to achieve results such as inflation, topology change or moduli stabilization, any interesting effects would be enclosed behind the horizon. Even very weak initial conditions ($|P|=1$) already contain apparent horizons before the simulations starts, and weaker conditions still form them within a small time. 
This shows that adding even a small initial momentum to appropriate metric functions ($g_{rr}$ and $g_{11}$, $g_{22}$ in this case) introduces a risk of creating black holes. A momentum which would solely change the size of the three sphere is disallowed by the Hamiltonial and momentum constraints, other changes to the inital conditions must be applied and the nature of these will determine the creation of a black hole.

\section{Conclusions}
We have used the static warped deformed conifold as an initial condition to which we have added a momentum tending to alter the size of the three sphere at the origin. The size of the three sphere seems very stable against momenta which would provoke it to shrink, this is explained by the stabilization of the sphere by the flux. The flux fixes the size of the three sphere and induces a potential that resists any change to the size. We have seen in the simulations that, due to the very steep potential in the direction of reducing the three-sphere, the sphere soon recovers its original size and, presumably, returns to the original geometry. The geometry was also seen to be remarkably robust to high-momentum perturbations in the direction of making the sphere smaller, with simulations even suggesting a minimum size of the three-sphere.

On the other hand, if we insert momentum which tends to increase the size of the three sphere then the growth slows down on a much longer timescale, as one would expect on physical grounds by diluting flux-lines rather than squeezing them. This leads to a shallow potential in this direction, and a geometry that is more susceptible to growth than collapse of the three sphere. We believe the growth would eventually stop and the size of the three sphere would tend back to its starting value, however this is not seen in simulation due to the much larger timescales involved.

The theme seems to be that the size of the three sphere cannot be collapsed to zero by adding initial momentum, the three sphere will reduce in size temporarily but will return to a value close to its starting size and will not vanish. This seems to rule out the possibility of forming a conical singularity when there are fluxes to consider. It does verify the stable radius of the three sphere in the Klebanov-Strassler static solution, as expected, and it also makes manifest the difficulty of dynamical topology changing transitions being realized in string theory when fluxes are present. We should, of course, make clear that there is a huge degree of freedom in choosing the initial data, despite the resrictions of the Hamiltonian and momentum constraint, and such conclusions are based on the initial data we chose.

While the size of the three-sphere may be stabilized by the fluxes, they do not guarantee anything about the risk of black hole formation. As we have demonstrated in our simulations, black holes can be formed from the throat with only a very small perturbation if it is in the "wrong" direction. This is a big risk (though not a certainty) in any procedure that involves dynamics on a deformed conifold, and such perturbations must be ruled out or otherwise addressed while making any models using probes on throats. The assumption that the manifold is unaffected by small probes may turn out to be too rash.

\acknowledgments
NB would like to thank STFC for financial support.

\bibliographystyle{unsrt}
\bibliography{throats}

\end{document}